\documentclass[aps,prl]{revtex4}
\usepackage{graphicx}
\def\ee{\end{equation}}
\def\bea{\begin{eqnarray}}

\def\Prob{{\rm Prob}}
\begin{document}

\title{Hodology}

\author{Adrian \surname{Kent}}
\email{A.P.A.Kent@damtp.cam.ac.uk} 
\affiliation{Centre for Quantum Information and Foundations, DAMTP, Centre for
  Mathematical Sciences, University of Cambridge, Wilberforce Road,
  Cambridge, CB3 0WA, U.K.}
\affiliation{Perimeter Institute for Theoretical Physics, 31 Caroline Street North, Waterloo, ON N2L 2Y5, Canada.}

\date{April 2020} 

\begin{abstract}
A {\it hodological} law 
causes the evolution of the universe to tend to follow
particular types of path.  
I give simple illustrations in toy models 
and discuss how Kolmogorov complexity
characterises the extent to which hodological laws 
explain, rather than merely describe, data.   
\end{abstract}
\maketitle
  
\section{Introduction}

If the probabilities we calculate in quantum theory are 
probabilities of some well-defined, objective, observer-independent features of nature, then
a complete formulation of quantum theory has to include a 
sample space on which these probabilities are defined.
The elements of that sample space form configurations
of beables, in Bell's terminology.
Whatever form they take, if they form part of physics as we understand
it they presumably have a mathematical structure.
It then makes sense to consider generalisations of quantum theory
in which the probabilities depend on that structure as well as 
the Born rule.    
 
This motivates looking at 
alternatives  \cite{kent1998beyond,kent2013beable} to cosmological
theories inspired by the standard understanding of quantum theory. 
Whatever the fine-grained form of the 
beables, generalised probability laws associated with them could also affect 
the probabilities of coarse-grained large-scale features of 
the universe.   The very large scale seems perhaps the most 
promising regime in which to look for empirical evidence 
of such deviations from quantum theory, since the strongest
evidence for quantum theory comes from small scale phenomena,
the relationship between quantum theory and gravity is not
known, and there are other outstanding cosmological puzzles
that suggest other lacunae in our understanding. 

Theories that are based on quantum theory but
guide the universe along paths other than those
implied by standard unitary quantum dynamics are not yet part of
standard mainstream discourse.
They impose constraints, in  a statistical sense.
However, these theories are qualitatively different
from quantum theory applied to constrained systems \cite{dirac2001lectures}. 
They are far more general than theories
with independent initial and final boundary
conditions.   
They are also far more general than dynamical collapse
models, although dynamical collapse models can be
seen as examples and can motivate others. 
Indeed, the generality they allow may raise a concern that
considering such theories takes us out of the domain of science: that they 
can describe data but cannot explain them in any standard scientific sense.   

I explain below why this concern is misplaced, using
simple models that show why these ``hodological''
theories are qualitatively different, 
illustrate their generality and explain the extent to which they
could nonetheless be scientifically useful. 

\section{Hodology in the Ehrenfest urn model}

The Ehrenfest urn model \cite{ehrenfesturn} nicely illustrates the 
effect of laws describing a statistical evolution from
an initial state.   It can be generalized to illustrate laws with independent
initial and final boundary conditions \cite{gell1994time}.  As we  
discuss below, it can also be generalized to illustrate hodological laws.  

The standard version of the Ehrenfest urn model begins with $N$ labelled balls
distributed between two urns ($A$ and $B$) in some initial
configuration (for example, all in urn $A$, or balls $1$ to
$\lfloor{\frac{N}{2}} \rfloor$ in $A$ and the rest in $B$). 
The model's state changes in discrete time steps, 
at each of which one label is chosen randomly, and the corresponding
ball switches urn.   
It is easy to see (analytically or numerically) 
that low entropy distributions typically evolve quickly towards
and then fluctuate around equipartition, spending nearly all the
time close to equipartition and returning to low entropy states
very infrequently. 

We consider the model with some number $T$ of time steps 
that is fixed in advance.  One might think of this as a toy model of a universe with
a fixed cosmological lifetime between its initial and final state. 
We take the numbers of balls $N_A$ and $N_B = N - N_A$ in urns $A, B$
to be the macro-physical variables of interest, and
the locations of each labelled ball to be micro-physical variables.  
Macrophysically, the possible evolutions from the initial state $N_A = N_A^0$ 
are thus given by sequences 
\begin{equation}
\underline{N_A} = ( N_A^0 , N_A^1 , \ldots , N_A^T ) \, , 
\end{equation} 
where $N_A^i = N_A^{i-1} \pm 1$.  
The sequence $\underline{N_A}$ has probability
\begin{equation}\label{urnprob}
\Prob( \underline{N_A} ) = \prod_{i=1}^T (  \delta (N_A^i - N_A^{i-1}, 1 ) \frac{ N - N_A^{i-1}
}{N} +
\delta (N_A^i - N_A^{i-1}, -1 ) \frac{ N_A^{i-1} }{N} ) \, .
\end{equation} 

We are interested in hodological generalisations that alter, and
are defined in terms of, the macrophysics.
To define such a model, we modify Eqn. (\ref{urnprob}), reweighting the
probabilities by non-negative factors $w (\underline{N_A} )$ that depend on the form of the path
$\underline{N_A}$ through configuration space.  
Thus 
\begin{equation}\label{urnprobmod}
\Prob_{\rm mod}( \underline{N_A} ) = C w (\underline{N_A} ) \Prob(
\underline{N_A} ) \, , 
\end{equation} 
where $C$ is a normalisation constant chosen so that 
\begin{equation}
\sum_{\underline{N_A}} \Prob_{\rm mod}( \underline{N_A} ) = 1 \, .
\end{equation}

\subsection{Simple examples}

\begin{figure}[h]
\centering
\includegraphics[width=0.5\linewidth]{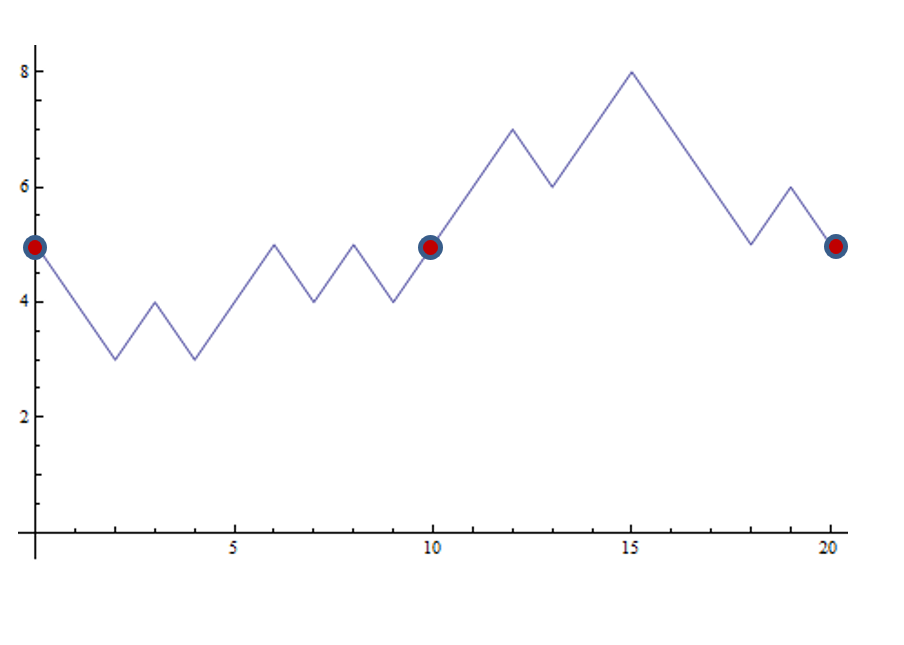} 
\caption{Single run of $N=10$ balls, constrained to $N_A = N_B =5$ at
  $t=0, 10$ and $20$. 
 }
\label{urn1}
\end{figure}

{\bf Example 1 (fixed macrophysical path points):} \qquad Let $N=10$, $T=20$, $N_A^0 = 5 $, and take  
\begin{equation}\label{mod1}
w(\underline{N_A} ) = \delta ( N_A^{10}, 5 ) \delta ( N_A^{20}, 5 ) \,
. 
\end{equation}
This weighting ensures that the realised evolution path has an equipartition
as its initial and final states and also at the midpoint of its
evolution.   A sample evolution is shown in Fig. \ref{urn1}.   

\begin{figure}[h]
\centering
\includegraphics[width=0.5\linewidth]{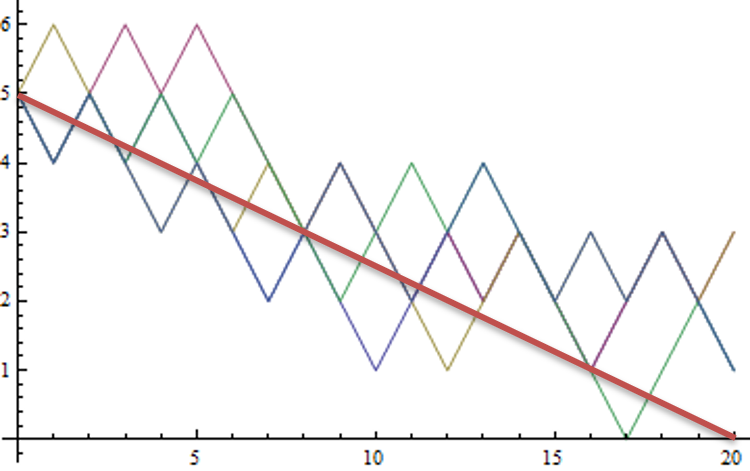} 
\caption{$5$ runs of $N=10$ balls, initial state $N_A = N_B =5 $ at
  $t=0$, drawn from an ensemble with evolution probabilities modified
  by the weight factor (\ref{mod2}).}
\label{urn2}
\end{figure}

{\bf Example 2 (weighting towards a given macrophysical path):} \qquad 
If, again with $N=10$, $T=20$, $N_A^0 = 5 $, we take  
\begin{equation}\label{mod2}
w(\underline{N_A}) = \exp( - \frac{1}{6} \sum_{t=1}^{20} ( N_A (t) - ( 5
- \frac{t}{4} ) )^2  ) \, , 
\end{equation} 
then the realised evolution path is likely to be relatively close to 
the line $N_A (t) = (5 - \frac{t}{4} ) $.   Sample evolutions are 
shown in Fig. \ref{urn2}.  

\subsection{Testing hodological laws}

Suppose now, for the sake of discussion, that we observe a new physical
system whose properties are opaque to us, except for one
discrete physical parameter that appears to evolve 
as though following some type of Ehrenfest urn model.
That is, there is one observable discrete parameter, $N_A$, which appears
always to lie in the range $0 \leq N_A \leq 10$. 
We observe its value only at discrete time steps $t$, after each of which
it increases or decreases by $1$.
Suppose we cannot measure anything else about the system  
(perhaps it is effectively a black box, or a very distant cosmological object
that regularly emits a discrete signal).    
Suppose also that, while we cannot directly observe 
the system's internal structure, extrapolating our knowledge of 
other better understood systems, and examining the evolution
statistics of $N_A$, lead us to the strong hypothesis that
it is characterised by some Ehrenfest model, 
with labelled subsystems playing roles analogous to 
those of the balls and urns.  
Suppose also that we have no information or
good hypothesis about any interaction with other systems.      
And suppose that the system goes through repeated runs of
$20$ time steps, apparently resetting (say after a gap of $10$ time
steps, so that individual runs are identifiable) after
each, with each run starting with $N_A=5$.  

After a while, we will conclude that, so long as we 
learn nothing more about the system, the only immediately scientifically
fruitful theories we can make about it are defined by generalised
Ehrenfest urn models of the form (\ref{urnprobmod}).
We can evaluate these by Bayesian
reasoning.  Informally, this would run roughly as follows. 
First, if our physical theories (the new system aside) 
take their current form, defined by initial states
and standard evolution laws, then before we examine the data
we would assign a high prior weight
to the standard Ehrenfest probability law (\ref{urnprob}),
i.e. to $C w( \underline{N_A} ) = 1$ for all paths
$\underline{N_A}$. 
We might assign a lower prior weight to the hypothesis that any modification of the
form (\ref{urnprobmod}) gives a better theory, and we would almost certainly assign
low prior weights to specific modified laws like (\ref{mod1}) and  
(\ref{mod2}).    But since the system is novel and mysterious, we 
should and probably would be undogmatic: every specific law $L$
would be assigned a non-zero prior weight $ \Prob_{\rm prior}(L)$.     

Suppose that on the first run we observed an evolution of the form of Fig. \ref{urn1}. 
According to the standard Ehrenfest probability law (\ref{urnprob}), 
the probability of equipartition of $10$ balls
at $t=10$, given initial equipartition, is 
$\frac{964533}{1953125} \approx \frac{1}{2}$. 
The probability of equipartition at both $t=10$ and $t=20$,
given initial equipartition, 
is thus $\approx \frac{1}{4}$. 

Bayesian hypothesis testing, given data $D$, assigns the 
posterior probability weight 
\begin{equation}
 \Prob_{\rm post}(L) = \frac{ \Prob ( D \, | \, L ) \Prob_{\rm prior}
   (L ) }{ \sum_i \Prob ( D \, | \, L_i ) \Prob_{\rm prior} (L_i ) }
 \, , 
\end{equation}
where the sum is over the set (which we assume countable) of all laws
considered. 

After the resulting Bayesian reweighting, our posterior weights 
for some of our modified laws would thus be smaller or zero, and our weight
for (\ref{mod1}) would (for sensible values of
$\Prob_{\rm prior} (L_i )$ ) be somewhat larger.
If our prior confidence in the law defined by Eqn. (\ref{urnprob}) was
high, our posterior confidence would still be high.   
However, if we saw $M$ runs, all of which produced evolutions with
equipartition at $t=10$ and $t=20$, the numerator in our posterior weight for (\ref{urnprob}) 
will be scaled by $( \frac{1}{4} )^M$, while the corresponding
expression for Eqn. (\ref{mod1}) remains unchanged.   
If the evolutions appear to be otherwise random, then 
our posterior weights for Eqn. (\ref{mod1}) increase with $M$,
tending to $1$ for large $M$.
In other words, we would eventually become very  
confident that the system is in fact governed by Eqn. (\ref{mod1}), 

Suppose instead that we saw an evolution of the type 
illustrated by Fig. \ref{urn2}. 
According to the standard Ehrenfest urn model, the probability of an
evolution as close as these to the line  $N_A (t) = (5 - \frac{t}{4} ) $ is roughly $1$ in $50000$.
Even after a single run, unless our prior weight for any law other
than (\ref{urnprob}) was significantly less than $ 2 \times 10^{-5}$, we would significantly lose confidence
in (\ref{urnprob}) and begin considering alternative laws seriously. 
After a small number of runs, we would likely
arrive at something like Eqn. (\ref{mod2}) as our best fit to 
the data.    

Since known physical
laws are based on probabilistic or deterministic evolution
from initial conditions, we might think a system 
apparently described by a modified Ehrenfest urn model such as
(\ref{mod1}) or (\ref{mod2})
must very likely have some additional internal mechanism and associated 
variables hidden from us, so that the complete system
is described by a more conventional law. 
We might then continue to search for ways of observing the 
hidden variables and obtaining a better and more detailed model.
Still, unless and until we succeeded, the relevant modified Ehrenfest urn model
would be our best description.   And we might not succeed: there need
not necessarily be any internal mechanism that gives any deeper
explanation. 
   
Formally, these calculations can be underpinned by the theory
of Solomonoff induction and the principle of minimum description
length (MDL) for hypothesis identification \cite{li2008introduction}.   
Roughly speaking, according to the MDL principle, 
the best hypothesis to fit the data is 
the one that minimizes the sum of the length of the program required  
to frame the hypothesis and the length of the string required to 
characterize the data given the hypothesis.        
The latter is approximately the Shannon entropy $S(H)$ of the probability
distribution on paths in variable space implied by the hypothesis $H$.
The former is the length $L(H)$ of a program mapping $\approx S(H)$ bit
strings to paths that, according to hypothesis $H$, are typical.
If $H_0$ is given by (\ref{urnprob}), 
$H_1$ by (\ref{mod1}) and $H_2$ by (\ref{mod2}), 
then for a single run
\begin{equation}
S(H_0 ) - S(H_1) \approx 2 \, , \qquad S(H_0) - S(H_2 ) \approx 16 \,
.
\end{equation}
For $M$ runs, $L(H_i )$ is fixed, while 
\begin{equation}
S(H_0 ) - S(H_1) \approx 2 M \, , \qquad S(H_0) - S(H_2 ) \approx 16 M \,
.
\end{equation}
Hence, if $H_1$ or $H_2$ fit the data, their description length becomes less than
that of $H_0$ for large $M$, and they become preferred to $H_0$; if no
more refined hypothesis fits the data then they become the MDL hypothesis.
The same is true of any hypothesis $H$ such that $S(H_0) - S(H) > 0$. 

\section{Discussion}

The Ehrenfest urn model illustrates how a model with
probabilistic microdynamics can be modified by 
macrodynamical laws that guide macroscopic variables
towards particular paths.   Such laws themselves may be either
deterministic (as in example (\ref{mod1})) or 
probabilistic (as in example (\ref{mod2})). 
It also illustrates how standard scientific inference
can identify such laws, if they offer a compressed
description of the observed data. 
 
Exactly the same points apply when we consider a 
microdynamics given by any version of quantum theory
that makes probabilistic predictions about the
microdynamics underpinning the physics of a macroscopic
system, including in principle the evolution of the universe.
As we have seen, one can model the evolution of a physical system
via an Ehrenfest urn model
without committing to identifying specific subsystems 
as balls and urns, or even committing to the belief
that such subsystems necessarily exist.  
Similarly, one can search for modified macrodynamical
laws in nature while remaining agnostic about precisely which events,
beables, or histories define the fundamental sample space for 
quantum theory.\footnote{
Examples of relevant versions of quantum theory include
theories with some form of
Copenhagen collapse rule,
quantum theory supplemented by mass-energy
beables determined mathematically by (fictitious)
asymptotically late time measurements,
a consistent history version of quantum theory 
defined via some appropriate set selection rule, 
an one-world version of quantum theory defined by
some appropriate selection rule for Everettian
branches, or some version of de Broglie-Bohm theory.}
The task is more complicated, because there are many
more possibly relevant variables and types of law.
Nonetheless, the methodology of Solomonoff induction applies. 
A systematic search for modified macrodynamical laws that might fit observation
better than standard quantum theory should be a major goal of
cosmological science, since in a strong sense it would necessarily advance
our knowledge.  Null results, excluding all
laws of a given type up to a given degree of complexity,
would solidify and parametrise our confidence in the
standard paradigm.   New laws would qualitatively change
our understanding of nature.

\section{Acknowledgements}
This work was supported by an FQXi grant and by Perimeter Institute
for Theoretical Physics. Research at Perimeter Institute is supported
by the Government of Canada through Industry Canada and by the
Province of Ontario through the Ministry of Research and Innovation.

\section*{References}
\bibliographystyle{unsrtnat}
\bibliography{qualia2}{}
\end{document}